\documentclass[conference]{IEEEtran}

\usepackage{cite}
\usepackage{amsmath,amssymb,amsfonts}
\usepackage{algorithmic}
\usepackage{graphicx}
\usepackage{textcomp}
\usepackage{comment}
\usepackage[nolist]{acronym}
\usepackage[nolist]{acronym}
\usepackage[normalem]{ulem}
\usepackage{comment}
\usepackage{color,soul}
\usepackage{multirow}
\usepackage{multicol,multirow,colortbl}
\usepackage[caption=false,font=footnotesize,labelfont=sf,textfont=sf]{subfig}
\usepackage{algorithm}
\usepackage[dvipsnames]{xcolor}
\usepackage{lmodern,textcomp}
\usepackage{adjustbox}

\UseRawInputEncoding

\begin{acronym} 
\acro{3GPP}{Third Generation Partnership Project}
\acro{5G}{fifth generation}
\acro{API}{application programming interface}
\acro{C-ITS}{cooperative-intelligent transport systems}
\acro{C-V2X}{cellular-vehicle-to-everything}
\acro{CAM}{cooperative awareness message}
\acro{CAV}{connected and autonomous vehicle}
\acro{CPM}{collective perception message}
\acro{DENM}{decentralized environmental notification message}
\acro{DSRC}{dedicated short range communication}
\acro{ETSI}{European Telecommunications Standards Institute}
\acro{FFT}{fast Fourier transform}
\acro{GNSS}{global navigation satellite system}
\acro{GPS}{global positioning system}
\acro{HAT}{hardware attached-on-top}
\acro{HMI}{human-machine interface}
\acro{IQ}{in-phase and in-quadrature}
\acro{ISAC}{integrated sensing and communication}
\acro{ITS}{intelligent transport system}
\acro{IVIM}{infrastructure to vehicle information message}
\acro{LDM}{local dynamic map}
\acro{LOS}{line-of-sight}
\acro{MAC}{medium access control}
\acro{MCM}{maneuver coordination message}
\acro{MIMO}{multiple input multiple output}
\acro{NLOS}{non-line-of-sight}
\acro{OBU}{on-board unit}
\acro{OFDM}{orthogonal frequency-division multiplexing}
\acro{OScar}{Open Stack for Car}
\acro{PHY}{physical}
\acro{RSU}{road side unit}
\acro{V2N}{vehicle-to-network}
\acro{V2I}{vehicle-to-infrastructure} 
\acro{V2V}{vehicle-to-vehicle} 
\acro{V2X}{vehicle-to-everything} 
\acro{VAM}{vulnerable road user awareness message}
\acro{WAVE}{wireless access in vehicular environment}
\end{acronym}

\begin{document}
\bstctlcite{IEEEexample:BSTcontrol}
%
\title{Low Cost C-ITS Stations Using Raspberry Pi and the Open Source Software OScar}

\author{\IEEEauthorblockN{
Lorenzo Farina\IEEEauthorrefmark{1}\IEEEauthorrefmark{2},
Matteo Piccoli\IEEEauthorrefmark{1}, 
Salvatore Iandolo\IEEEauthorrefmark{3},
Antonio Solida\IEEEauthorrefmark{3},
Carlo Augusto Grazia\IEEEauthorrefmark{3},\\
Francesco Raviglione\IEEEauthorrefmark{4},
Claudio Casetti\IEEEauthorrefmark{5},
and Alessandro Bazzi\IEEEauthorrefmark{1}\IEEEauthorrefmark{2}
}\\
\IEEEauthorblockA{\IEEEauthorrefmark{1}DEI, Universit\`a di Bologna, 40136 Bologna, Italy}
\IEEEauthorblockA{\IEEEauthorrefmark{2}National Laboratory of Wireless Communications (WiLab), CNIT, 40136 Bologna, Italy}
\IEEEauthorblockA{\IEEEauthorrefmark{3}DIEF, University of Modena and Reggio Emilia, 41125 Modena, Italy}
\IEEEauthorblockA{\IEEEauthorrefmark{4}Dipartimento di Elettronica e Telecomunicazioni, Politecnico di Torino, 10129 Torino, Italy}
\IEEEauthorblockA{\IEEEauthorrefmark{5}Dipartimento di Automatica e Informatica, Politecnico di Torino, 10129 Torino, Italy}
}


\markboth{Journal of \LaTeX\ Class Files,~Vol.~14, No.~8, August~2015}%
{Shell \MakeLowercase{\textit{et al.}}: Bare Demo of IEEEtran.cls for IEEE Transactions on Magnetics Journals}
%





\IEEEtitleabstractindextext{%
\begin{abstract} 
The deployment of \ac{C-ITS} has started, and standardization and research activities are moving forward to improve road safety and vehicular efficiency. An aspect that is still felt as a limitation by the research groups active in the field, is the difficulty to validate the solutions with real hardware and software, because of the huge investments that are needed when multiple equipped vehicles need to be considered. In this work, we present a platform with low-cost hardware based on a Raspberry Pi and a Wi-Fi module transmitting at 5.9~GHz, and on the open-source software \ac{OScar}, which is compliant with the ETSI \ac{C-ITS} standards. With a limited cost in the order of 200~\texteuro, the platform realizes a device which is standard compliant and can be used as either \ac{OBU} or \ac{RSU}. The limited cost makes the testbed scalable to several units with limited budget and the limited size makes it also deployable on mini-cars to test advanced \ac{CAV} networks and applications. Our tests demonstrate its interoperability with other devices, compliance in terms of power spectrum, and a range of a few hundred meters in \ac{LOS} conditions using the standard settings of ITS-G5.
\end{abstract}
}

\maketitle

\IEEEdisplaynontitleabstractindextext

%
\IEEEpeerreviewmaketitle

\acresetall

\section{Introduction}

After decades of studies and standardization effort, the deployment of cooperative \ac{V2X} devices on vehicles and roads have eventually begun.\footnote{In Europe, some models started to be equipped by default with OBUs in early 2021, and several initiatives like C-ROADS deployed RSUs in many european highways and cities. It is estimated that there are today at least 1.5 million cars equipped and tens of hundreds of kilometers covered by C-ITS.} Even if the number of equipped cars is still limited, effort is ongoing to move from basic to advanced services and to increase the performance and impact of \ac{V2X} on road safety and efficiency. 

\begin{figure}[t]
\centering
\includegraphics[width=80mm]{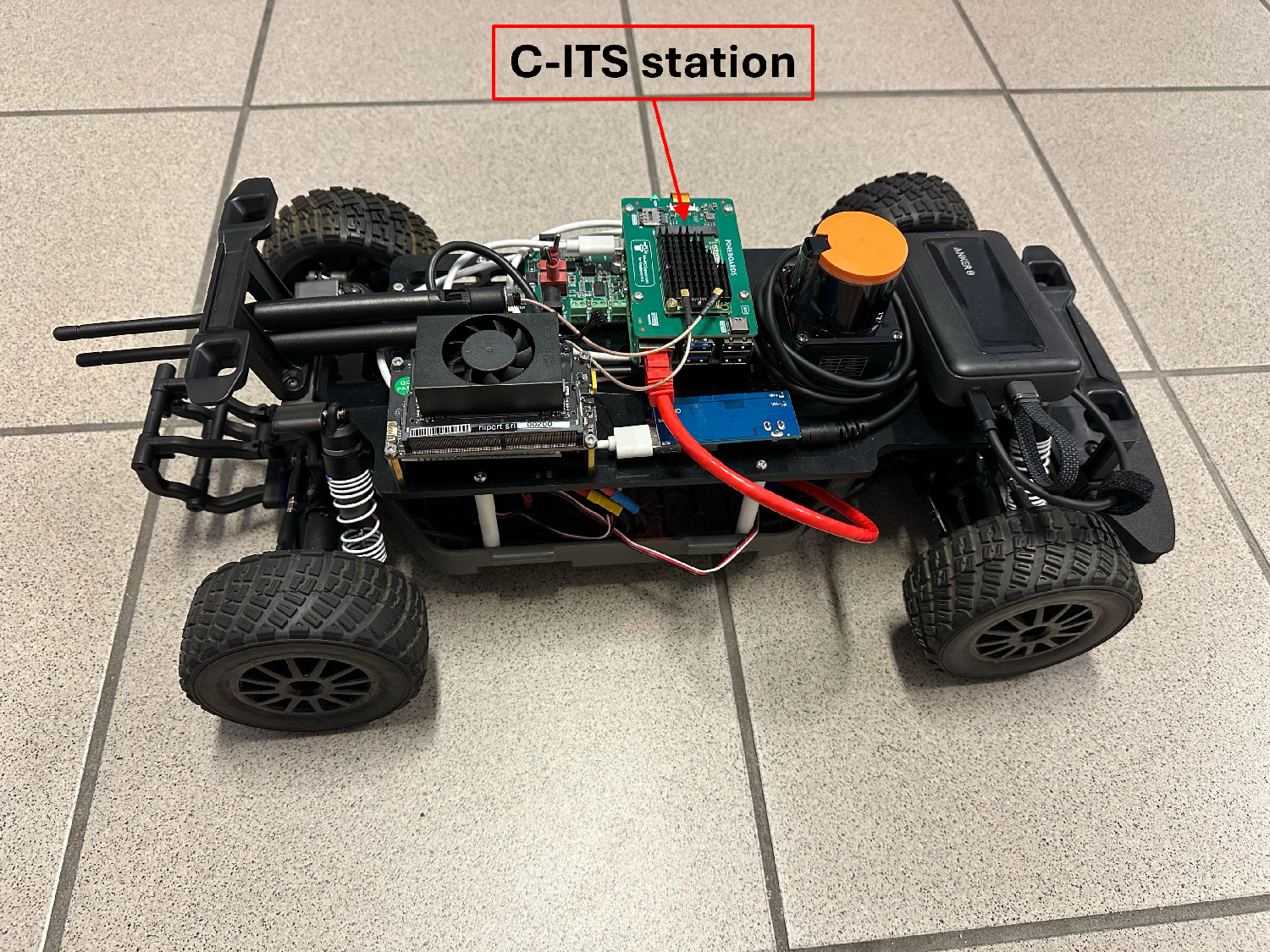}
\caption{Mini-car scale 1:10, equipped with the low-cost OBU.} 
\label{fig:f1tenth}
\end{figure}


From a standardization perspective, in Europe ETSI has completed the definition of a set of standards indicated as the Release~1 of \ac{C-ITS}. Most of the data exchange in Release~1 is based on \acp{CAM} and \acp{DENM}, which broadcast information about the vehicle status and mobility, or about particular events like the presence of road works or the approaching of an ambulance. The work is currently continuing towards the C-ITS Release~2 \cite{10467184}, which extends the awareness to other road users with \acp{VAM} and allows them to exchange what is observed through \acp{CPM}. Work is also ongoing to enable vehicles to agree on their operations for maneuver management through \acp{MCM}.


One of the main challenges for the research groups active in this field is how to move from simulations to field trials. The cost of a single device is in fact often in the order of a few thousand euros, which makes the realization of testbeds with more than a few units very expensive. If it is hardly affordable when looking only at the devices used as hardware-in-the-loop, it becomes prohibitive when devices need to be mounted on vehicles moving on the road.

To cope with this limitation, we have realized a low-cost platform that can act as \ac{OBU} or \ac{RSU}, mounting an open-source fully compliant \ac{C-ITS} stack. More specifically, the hardware is based on the Raspberry~Pi~5 and the software on the \ac{OScar} framework. A single device has a cost in the order of  200~\texteuro, making the testbed scalable even with a limited budget. The size is also very limited, since it is smaller than a normal smartphone (although a bit thicker). The low cost and reduced size make it possible to realize \acp{CAV} starting from scaled versions of real vehicles, as shown in Fig.~\ref{fig:f1tenth} using F1tenth mini-cars \cite{burgio2, okelly2019f110opensourceautonomouscyberphysical}. It should be remarked that the proposed platform does not have the goal to replace devices with higher cost, which can clearly guarantee higher performance, but to create an intermediate step between simulations and products ready for the market. 

In this paper, we provide a summary of a measurement campaign we performed to validate the platform, which showcases its performance in terms of spectrum usage, power, and range. In the following, we first provide some details on the hardware and software used  in Section~\ref{sec:station}, and then discuss our measurements in Section~\ref{sec:measurements}. Our conclusions are finally drawn in Section~\ref{sec:conclusion}.

\section{Details of the low-cost C-ITS station}\label{sec:station}

In this section, some details about the low-cost hardware and open-source software are provided.

\subsection{Hardware}

\begin{figure}[t]
\centering
\includegraphics[width=80mm]{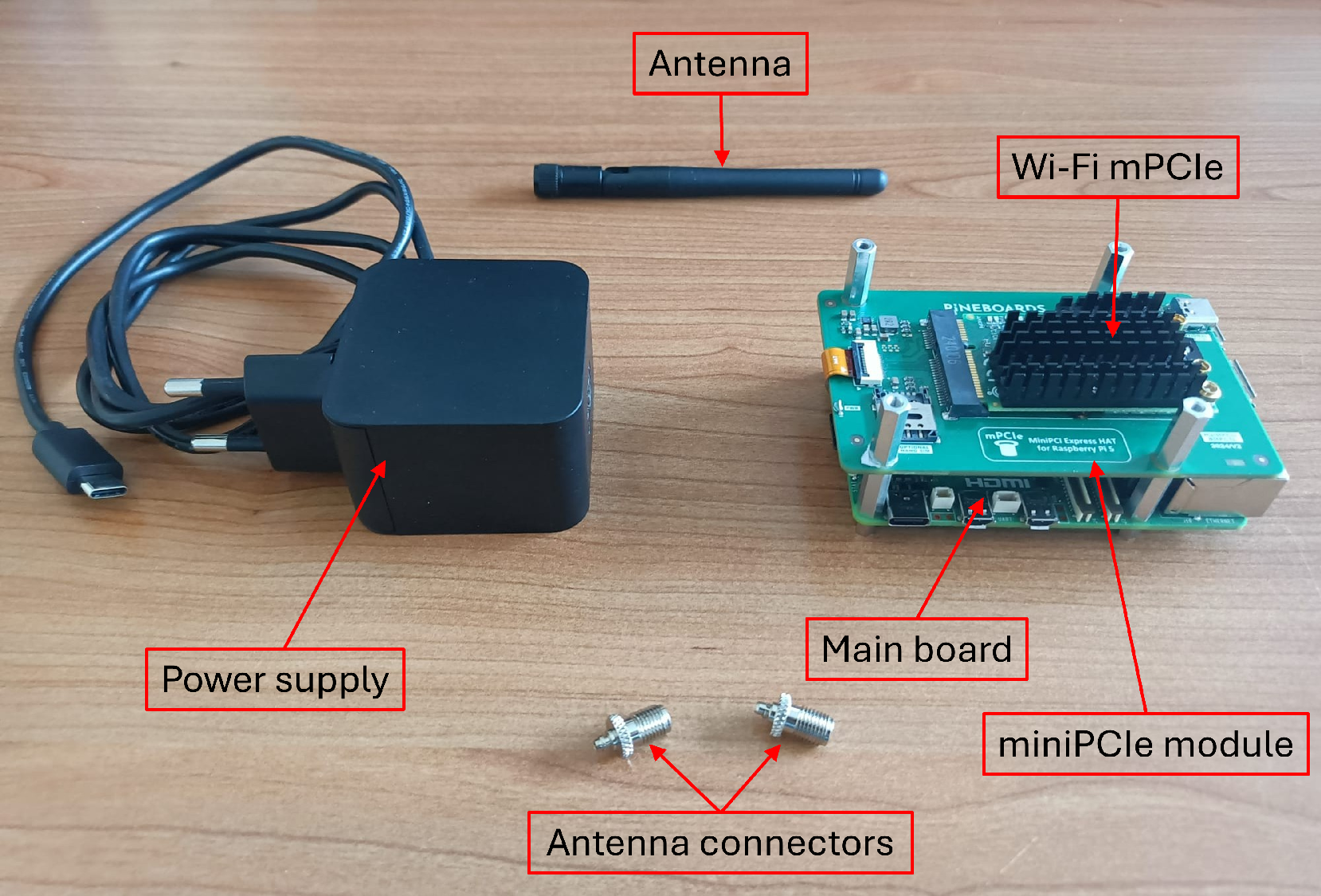}
\caption{The hardware components.} 
\label{fig:hwelements}
\end{figure}

A detailed list of the components used to realize the device is provided in  Table~\ref{tab:hwlist} and illustrated in Fig.~\ref{fig:hwelements}. The cost order of each component is also provided. The most relevant parts in terms of cost are the Raspberry Pi, the Wi-Fi mPCIe, and the power supply.

The Raspberry Pi 5 Model B is a compact single-board computer featuring an ARM Cortex-A76 processor, 8 GB of LPDDR4 RAM, and 128 GB of flash storage. It is powered via a 5V/5A DC USB-C connection and includes a PCIe 2.0 interface for high-speed peripheral expansion. This flexibility enables the use of various \ac{HAT} add-ons, following a standardized format, such as M.2 SSDs for improved read/write performance. In our setup, we integrated a PCIe HAT to accommodate a Wi-Fi card that meets our specific needs. We chose the Mikrotik R11E-5HND, an IEEE~802.11a/n adapter built on the Qualcomm Atheros AR9580 chipset and designed with a thermal heat sink for efficient cooling. This chipset operates in the upper 5 GHz spectrum and is compatible with the ath9k Linux kernel driver, which can be modified to enable V2X communication using the IEEE 802.11p standard \cite{grazia2018performance}. A minor point of attention is that the Raspberry Pi 5 requires 25 or more Watts from the power source, therefore an appropriate power supply needs to be used; this is to be taken into account especially if a battery (e.g. a power bank) is used.

\subsection{Software}

The main software component of our platform is represented by the OScar framework, a lightweight, open-source, self-contained implementation of the ETSI \ac{C-ITS} stack, with the related basic services, designed to run on Linux \cite{OScar_paper_2024}. OScar has been developed in C++, keeping the number of required dependencies as little as possible, to enable its integration not only on more powerful \acp{OBU} and on development hardware, but also on low-cost and low-power devices running dedicated Linux distributions such as OpenWrt \cite{OpenWrt_V2X_paper_2019} or Raspberry Pi OS. Furthermore, it follows a modular and multi-threaded approach to take advantage, if available, of multi-core CPUs. The whole framework can anyway run on a single core, while still guaranteeing a good performance, with each thread managing a different set of operations. More specifically, OScar is organized in: (i) one thread for reception of ETSI \mbox{C-ITS} messages, (ii) a set of threads each managing the transmission of a different message type, (iii) a thread that manages a web-based \ac{HMI}, (iv) a thread that periodically checks the consistency of the internal \ac{LDM} and clears old entries, (v) a thread in charge of managing the \ac{API} to retrieve information from the \ac{LDM}, and, finally, (vi) a thread that is optionally spawned every time OScar needs to read information from the vehicle CAN bus.

    	\begin{table}[t]
	\caption{List of hardware components.}\label{tab:hwlist}
		\centering 
        \scriptsize
	\begin{tabular}{m{2.2cm}p{4.4cm}p{0.9cm}}
\hline \hline
\textbf{Device} & \textbf{Details} & \textbf{Cost} \\ \hline 
 Main board & Rasberry Pi 5 Model B & 80~\texteuro \\ 
Power supply & Cable-supply or power bank, $>$25~W & 15-25~\texteuro \\ 
Memory & MicroSD, at least 32~GB & 	10~\texteuro \\ 
miniPCIe module & Rasp5 HAT for miniPCIe & 15~\texteuro \\ 
Fasteners & Spacers (4) and screws (4) & 5~\texteuro \\ 
Wi-Fi mPCIe & R11E-5HND ath9k & 25~\texteuro \\ 
Antenna conn. & From MMCX to SMA & 5~\texteuro \\ 
Antenna & SMA antenna for 5.9~GHz & 5~\texteuro \\ 
\hline \hline
\end{tabular}
	\end{table}

\begin{figure*}[t]
\subfloat[]{\includegraphics[width=0.65\columnwidth]{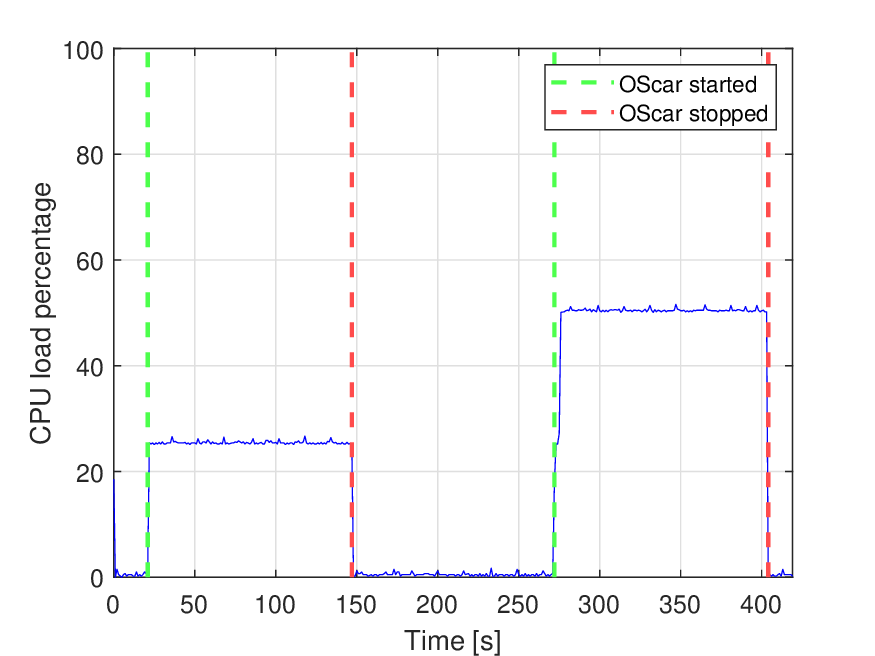}
\label{fig:oscarimpact_cpu}}
\hfill
\subfloat[]{\includegraphics[width=0.65\columnwidth]{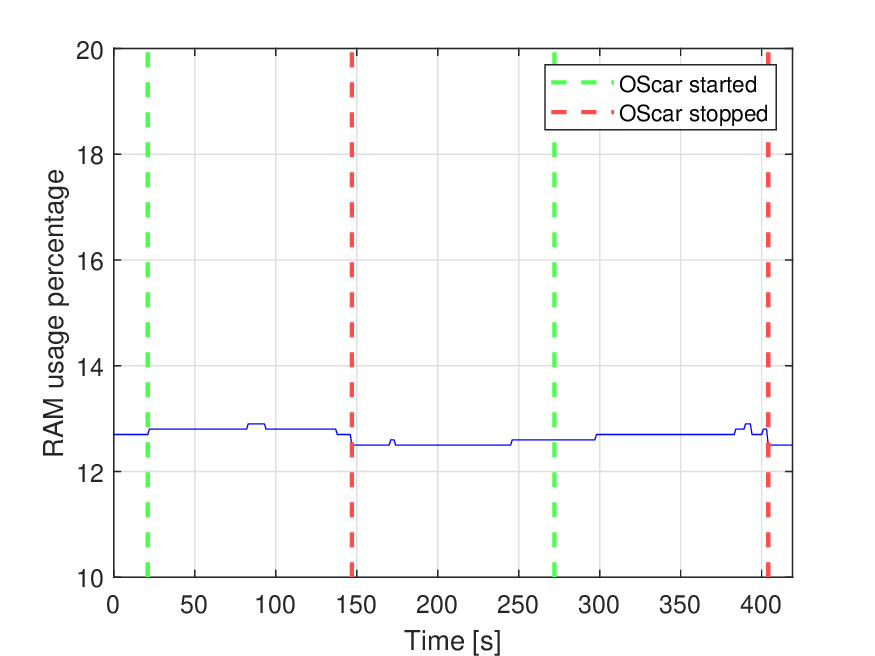}
\label{fig:oscarimpact_ram}}
\hfill
\subfloat[]{\includegraphics[width=0.65\columnwidth]{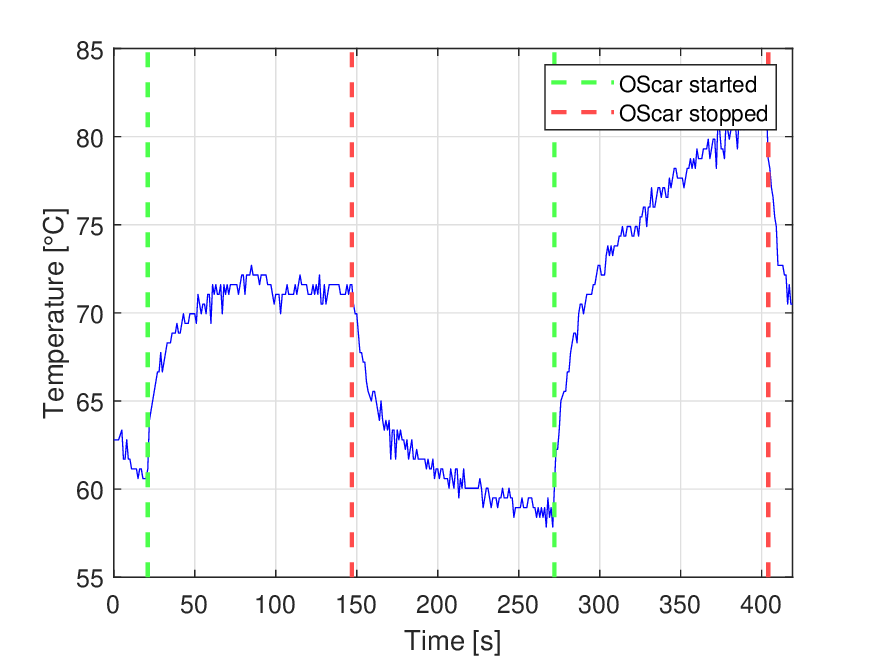}
\label{fig:oscarimpact_temp}}
\caption{Impact of OScar on the Raspberry Pi. (a) CPU. (b) RAM. (c) Temperature.}
\label{fig:oscarimpact}
\end{figure*}

OScar currently includes the full implementation of the most relevant ETSI safety-related services, supporting the transmission and reception of \acp{CAM}, \acp{VAM}, and \acp{CPM} (being the first framework to support the recently standardized version 2), both unsecured and with L0 security,\footnote{L0 is currently the level of security certification that is provided for research purposes by the European Public-Key Infrastructure} as foreseen by the ETSI standards. It also embeds a highly efficient \ac{LDM}, storing the information about perceived and received objects and road users, that can be queried by external services thanks to a dedicated JSON-over-TCP \ac{API}. In addition, it is currently being extended to support additional security functionalities, besides event-triggered messages such as \acp{DENM} and \acp{IVIM}.

Finally, OScar is able to retrieve positioning and sensor data, required for the transmission of standardized messages, from different sources depending on the target platform on which it is deployed. As an example, \ac{GNSS} positioning data can be obtained by connecting to the standard \texttt{gpsd} Linux daemon, or by directly parsing the stream of bytes from the serial device corresponding to the GNSS receiver, which may also include an embedded inertial sensor. In this second case, OScar relies on an additional thread that manages the parsing of the information from the serial device, and its provision to the other ETSI services running on separate threads.

\section{Experimental Results}\label{sec:measurements}

The platform has been tested through both laboratory and on-field experiments. In order to assess the performance of the Raspberry Pi when using \ac{OScar} to act as a C-ITS station, we focused on four key aspects, namely the impact of OScar on the resources of the Raspberry Pi, the transmitter spectrum, the transmitted power, and the communication range. 

\subsection{Impact of OScar on the Raspberry Pi}
The first measurements performed aimed to assess the impact of \ac{OScar} on the Raspberry Pi in terms of CPU load, RAM usage, and temperature.

To this end, a Python script recording the three metrics over time was executed in the background. While the script was running, \ac{OScar} was launched and stopped twice, each time running for approximately two minutes. The first time, the software was started enabling the reception only, while the second time CAMs were generated too. In both cases, the \ac{LDM} was created and the software was connected to the \texttt{gpsd} Linux daemon. The transmitted messages were unsecured. 

The results are depicted in Fig.~\ref{fig:oscarimpact}, with the three subfigures showing the time on the x-axis and the plotted metric on the y-axis. 
Fig.~\ref{fig:oscarimpact_cpu}, in particular, shows the CPU load. It can be noted that the impact of \ac{OScar} on the CPU is not negligible, with the software requiring 25\% of the CPU capacity to run and an additional 25\% to generate CAMs. On the contrary, the impact on the RAM usage is light, as highlighted in Fig.~\ref{fig:oscarimpact_ram}. Finally, Fig.~\ref{fig:oscarimpact_temp} shows that the temperature increases from a baseline value of around 60$^\circ$C
to approximately 80$^\circ$C when transmitting CAMs, suggesting that an additional cooling system on the Raspberry Pi would be beneficial.

\subsection{Signal spectrum}\label{sec:measurements_spectrum}
Among the requirements for a C-ITS station to be standard-compliant, one concerns the transmitter spectrum. In particular, ETSI EN 302 571~\cite{etsi-302571} defines a transmitter spectrum mask that should not be exceeded when transmitting with a channel bandwidth of 10 MHz.

In order to verify whether the implemented platform satisfies this requirement, CAMs were transmitted at the maximum available power of 26~dBm and received by a spectrum analyzer connected to the device through an SMA cable. The employed spectrum analyzer for this part of the campaign is a Rohde \& Schwarz FPL 1007.

Since CAMs are generated at a maximum frequency of 10 Hz and not continuously, properly observing the spectrum is not straightforward. Our approach was to first capture the received signal over a time interval of 1 second, which is a built-in feature of the instrument, and then retrospectively  filter out the intervals where only noise was present at the receiver. 
More specifically, the spectrum analyzer was first used to sample the received signal $x(t) = p(t) + j q(t)$, where $p(t)$ and $q(t)$ are the baseband components in the phase and quadrature, respectively. The obtained samples were then processed in Matlab. 
During our measurement, a bandwidth of 12.8~MHz, which is the maximum allowed by the spectrum analyzer, was acquired around the central frequency of 5.9 GHz over a time interval of 1 second. Then, the samples were exported on Matlab, where intervals without useful signal were removed. From a practical perspective, this means that the samples with a power below a minimum threshold are interpreted as noise and discarded. 
Next, the \ac{FFT} was applied to the remaining samples 
to obtain the frequency-domain representation of the signal. More specifically, denoting as $x_i$ the sample of $x(t)$ at time $t_i$, $\bold{x}_i=\{x_i,x_{i+1},...,x_{i+N_\text{F}-1}\}$ a vector of $N_\text{F}$ consecutive samples from the instant $t_i$, $\bold{X}_j=\textit{F}\{\bold{x}_i\}$ the FFT of $\bold{x}_i$, and $X_j$ the output of the \ac{FFT} at frequency $f_j$, the power spectral density in dBm/Hz was computed at the same frequencies as $P(f_j) = 10\cdot \log_{10}\left(\frac{|X_j|^2}{R\cdot b\cdot N_\text{F}}\right)+30$, where $R$ is the system impedance and $b$ is the width of the frequency bin associated to a single frequency sample.

\begin{figure}[t]
\centering
\includegraphics[width=\columnwidth]{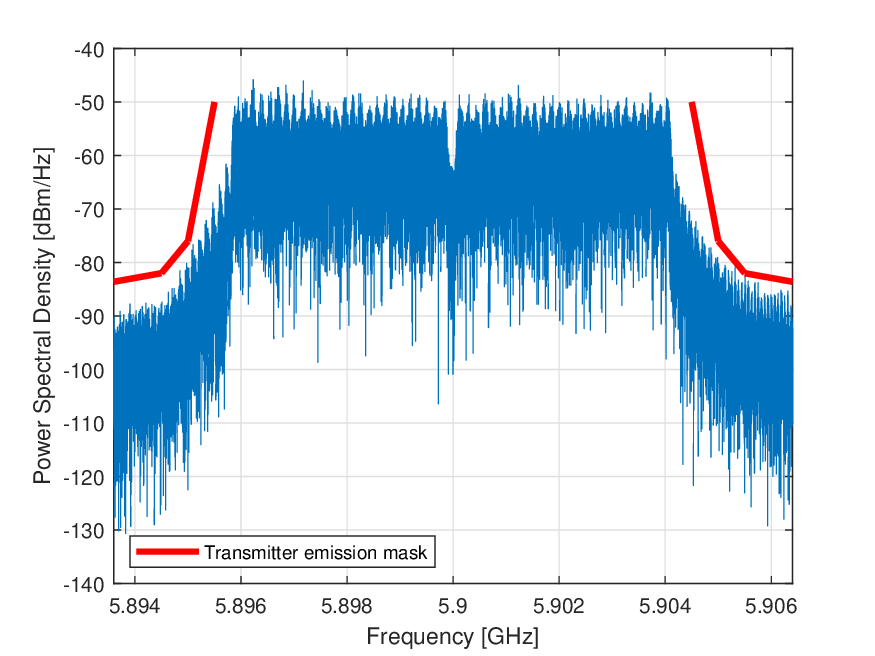}
\caption{Spectrum of the signal with emission mask defined in ETSI EN 302 571~\cite{etsi-302571}.}
\label{fig:spectrum}
\end{figure}

The resulting power spectral density, , obtained using $N_\text{F}=65536$, is shown in Fig.~\ref{fig:spectrum}, along with the emission mask limited to the acquired bandwidth. 
As observable, the spectrum remains below the limits over the whole analyzed bandwidth, as requested by the standard.

\subsection{Characterization of the transmitting power}


Another aspect to explore is what power is actually obtained in output when a certain power level is set; additionally, it is helpful to verify that a linear variation is obtained to the output power when the set power is modified.



%

\begin{figure}[!t]
\centering
\subfloat[]{\includegraphics[width=0.8\columnwidth,draft=false]{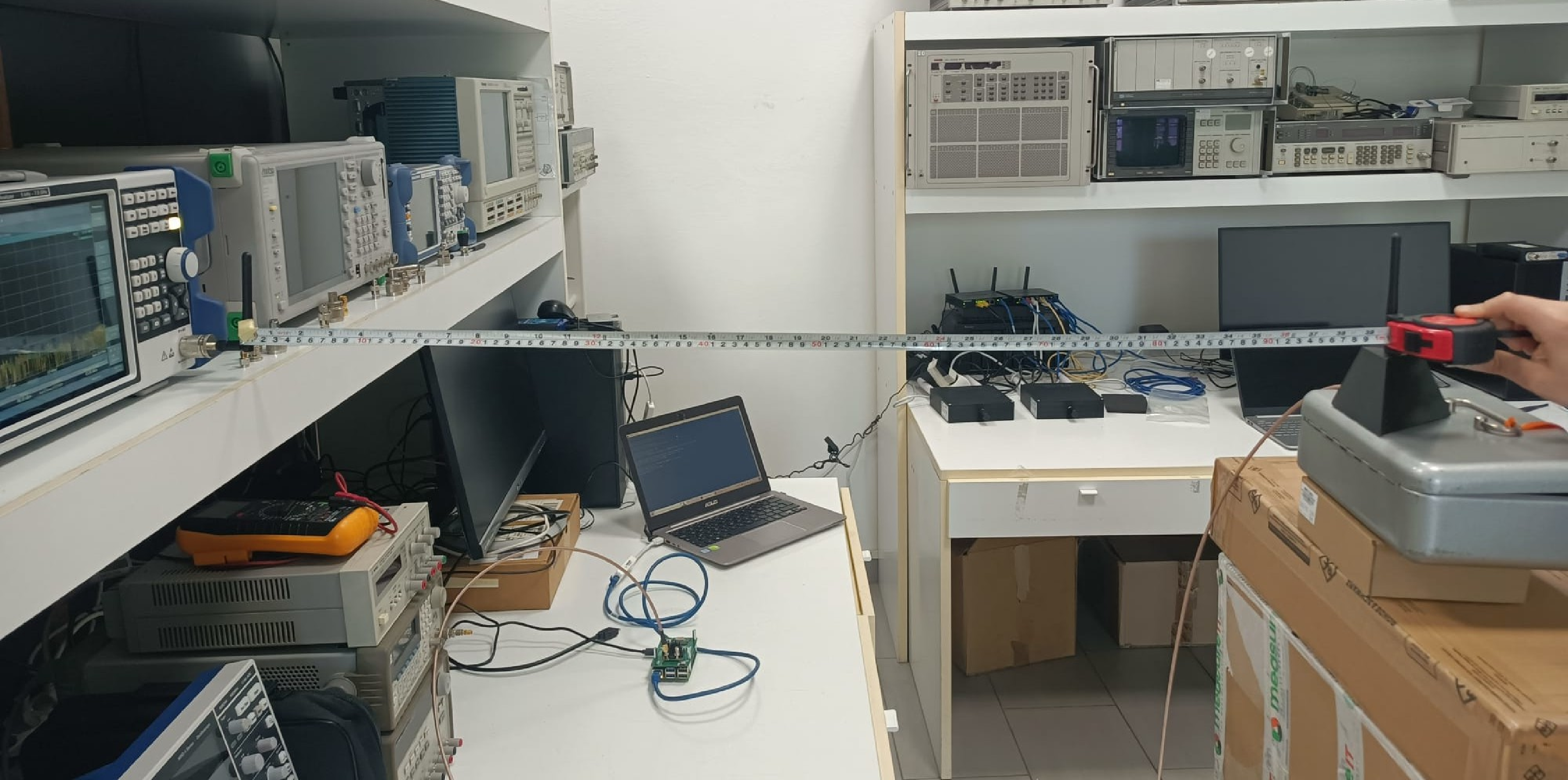}
\label{fig:powerlevels_setup}}\\
\subfloat[]{\includegraphics[width=\columnwidth,draft=false]{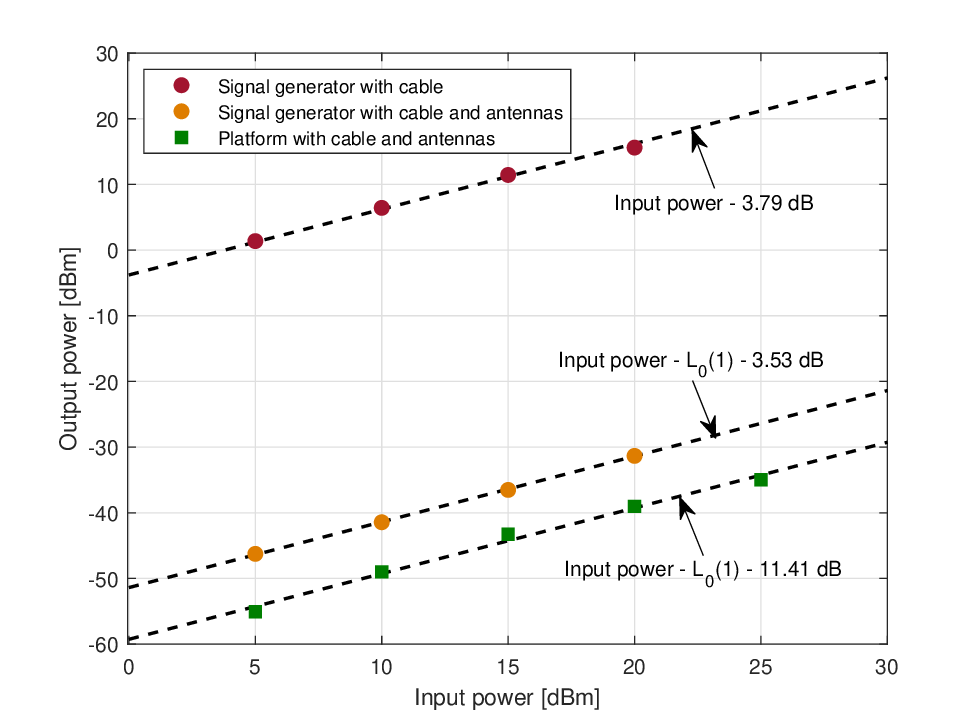}
\label{fig:powerlevels_curves}}
\caption{Power related measurements. (a) Setup. (b) Results.}
\label{fig:powerlevels}
\end{figure}

The power characterization was performed through laboratory measurements, using the same spectrum analyzer detailed in Section~\ref{sec:measurements_spectrum}. 
%
%
In addition to the use of the implemented platform to generate the signal, measurements were carried out also using a signal generator, which continuously transmitted a sinusoid at a frequency of 5.9 GHz. This approach enables a more reliable characterization of the SMA cable and provides a benchmark for the platform's power characterization. The employed signal generator is an Agilent 83650B.

The aim was eventually to measure the average received power by the spectrum analyzer when the platform is transmitting CAMs, varying the transmitted power levels. Both instruments were equipped with an antenna,\footnote{The antennas used are from Molex for ISM/DSRC 5.8-6 GHz, with nominal gain 3.9 dBi.} and the distance between the transmitting and receiving antenna was 1 meter (as shown in Fig.~\ref{fig:powerlevels_setup}). In order to ensure that the distance remained constant throughout the measurements, the Wi-Fi module was connected to the antenna through the SMA cable, and a stabilizing antenna support was used. 
Measurements with the signal generator were performed in the same experimental setup, as well as when the signal generator was directly connected to the spectrum analyzer through the SMA cable. 
As described in the previous subsection, the spectrum analyzer was used also in this case to sample the received signal and obtain the samples $x_i$; when the signal was generated by the proposed platform, the samples were also elaborated in Matlab to remove noise intervals. Then, the average received power in dBm was computed as $\overline{P} = 10\cdot \log_{10}\left(\frac{\sum_{i=1}^{N_\text{s}} |x_i|^2}{N_\text{s}\cdot R}\right)+30$, where $R$ is the system impedance and $N_\text{s}$ is the number of samples available case by case. 

\begin{figure}[t]
\centering
\includegraphics[width=\columnwidth,draft=false]{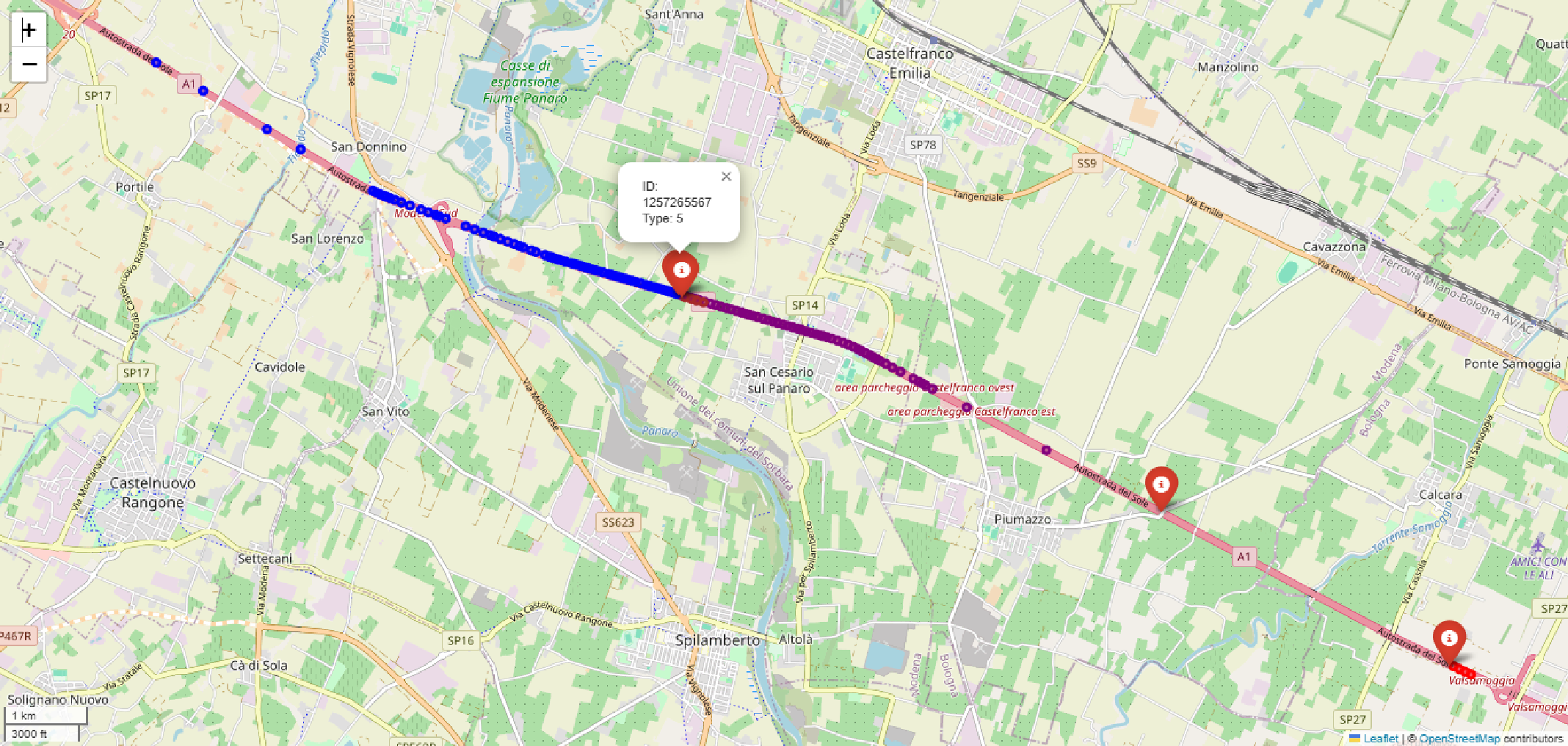}
\caption{Representation of the CAMs captured driving on a highway. Each circle represents the average position of ten messages received. Different colors correspond to different vehicle IDs; per each of them, a red marker indicates where the first CAM was received.}
\label{fig:highwaycapture}
\end{figure}

Measurement results with the signal generator and our platform are shown in Fig.~\ref{fig:powerlevels_curves}. The input power in dBm is depicted on the x-axis, while the output power in dBm is represented on the y-axis. The output power when transmitting with our platform was recorded for five different values of input power, while only four values were considered with the signal generator, as the instrument is not able to reach a power of 25 dBm in transmission.
Each set of measurements is represented in the figure by a group of markers, with a dashed line associated to each group to highlight the average offset between the input and output powers. This offset is the attenuation in dB introduced by the cable and connectors, to which the free-space loss 
has to be added when using the antennas. The free-space loss in dB is computed as $L_0(d) = 20\cdot \log_{10}\left(\frac{4\cdot\pi\cdot d \cdot f_c}{c}\right)$, where $d$ is the distance between the antennas, $f_c$ is the carrier frequency, and $c$ is the speed of light.

Observing the results obtained with the signal generator, it is evident that the attenuation caused by the cable and connectors remains almost constant, either transmitting with or without the antenna; this also implies that the gain due to the antennas is counter-balanced by the loss of the antenna connections. When transmitting with the platform, the attenuation experienced before the antenna seems to increase, and this is probably caused by the connectors used to link the Wi-Fi module to the SMA cable. Nevertheless, the output power clearly shows a linear increase with increasing input power, demonstrating that the device is able to properly adjust the transmitted power as requested.


\begin{figure}[!t]
\centering
\subfloat[]{\includegraphics[width=0.8\columnwidth,draft=false]{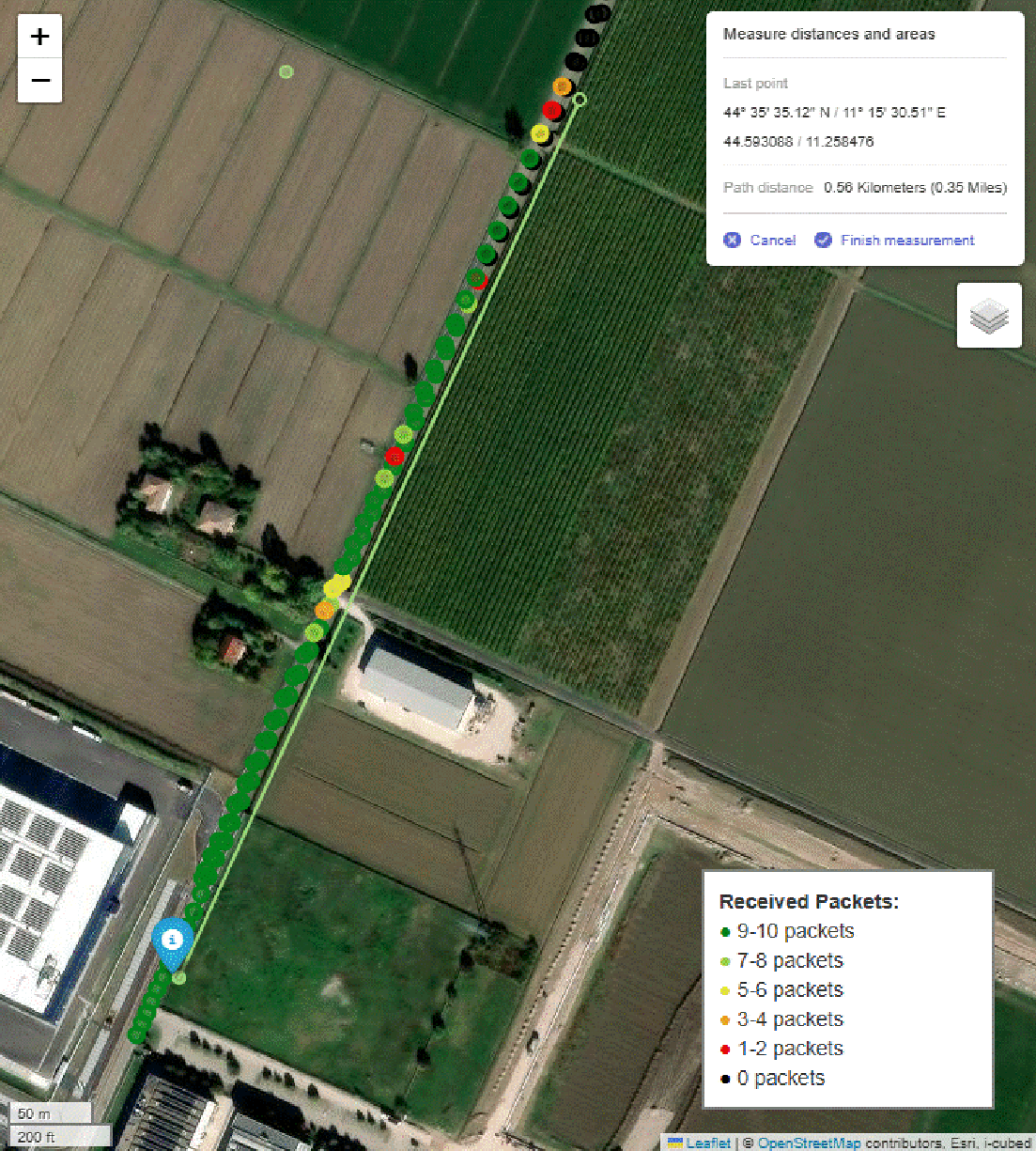}
\label{fig:range_los}}\\
\subfloat[]{\includegraphics[width=0.8\columnwidth,draft=false]{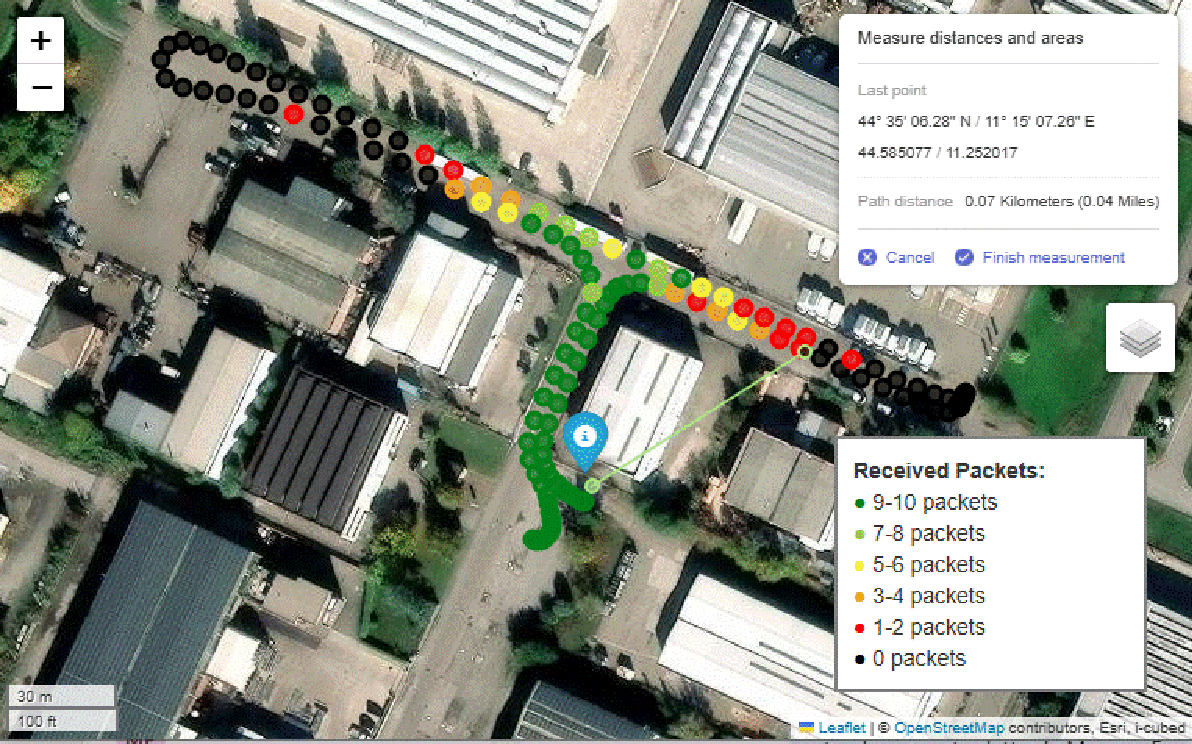}
\label{fig:range_nlos}}
\caption{Tests from field trial. Each circle corresponds to ten CAMs transmitted: the position on the map is the average location, while the color indicates the number of correctly decoded CAMs. (a) LOS. (b) NLOS.}
\label{fig:range}
\end{figure}

\subsection{Field trial}

Lastly, on-field measurements were carried out to verify the correct operations of the platform and to estimate its communication range. 

A first test was carried out with one Raspberry Pi acting as receiver, mounted on a car driving in a highway. The Raspberry Pi was placed inside the car and powered by an inverter connected to the car's cigarette lighter. Moreover, an antenna and a GNSS module\footnote{The localization was obtained using the GPS module VK-162, with update rate set to 10~Hz.} were placed on the car roof, with the first linked to the Raspberry Pi through an SMA cable and the second connected directly through USB. This test allowed us to verify that the device is able to correctly receive and decode messages from other \mbox{C-ITS} stations.
During our test, we received CAMs, DENMs, and IVIMs from 8 RSUs and 9 OBUs. An example of our observations is shown in Fig.~\ref{fig:highwaycapture}, where the CAMs received over approximately 30~km are shown; in the figure, each color represents a different ID of the transmitting station, and each circle is located at the average position indicated by ten consecutive received CAMs; the red marker shows the position where the first CAM was received per each ID. 

In addition to this verification, we also performed a set of range measurements using two Raspberry Pis. 
One of the two cars remained stationary for the duration of the measurements and transmitted one CAM every 100~ms\footnote{For the purpose of increasing the number of transmitted messages, in these experiments we forced the generation of 10 CAMs per second,  ignoring the triggering conditions imposed by ETSI; the triggering conditions are anyway correctly implemented in OScar, as detailed in \cite{OScar_paper_2024}.} with a transmit power level set to 26~dBm; the other car followed a predefined path while recording the received messages. A number of different tests were performed, including scenarios without large obstacles, hereafter called in \ac{LOS} conditions, and others where buildings were obstructing the communications along portions of the path, called in \ac{NLOS} conditions. 

The results obtained in the different tests were similar. Those obtained in one scenario in \ac{LOS} conditions and in one in \ac{NLOS} conditions are shown in Fig.~\ref{fig:range}. The blue marker indicates the position of the stationary car. Each plotted point is the average position of the moving car over a one-second time window, with colors reflecting the number of messages that are correctly decoded in that second. Since the stationary car transmits one message every 100~ms, the number of messages received in one second can vary from a minimum of 0 (represented in black) to a maximum of 10 (represented in dark green).

Observing Fig.~\ref{fig:range}, a maximum communication range of around 560 meters is obtained in the scenario in LOS conditions (Fig.~\ref{fig:range_los}), while approximately 70 meters were reached behind a building in the scenario in NLOS conditions (Fig.~\ref{fig:range_nlos}).  Looking at the LOS scenario, it can be noted that at a distance of about 250~m more packets than at larger distances were lost; this is to be attributed to the effect of the reflection on the road, which creates a two-ray propagation with losses of received power at some specific distances, coherently with what has been shown for example in \cite{HASIRCITUGCU2024100791,10.1145/2482967.2482980}. Overall, these distances appear lower than those obtained with more costly devices, but still acceptable for the testing purposes targeted by our platform.

\section{Conclusion}\label{sec:conclusion}
In this paper, we discussed the realization and testing of a low cost platform acting as a C-ITS station, based on a Raspberry Pi 5 and the open-source software \ac{OScar}. The performance of the device has been evaluated focusing on the impact of \ac{OScar} on the Raspberry Pi, the transmitter spectrum, the transmitted power, and the communication range. Measurements have been performed both in a laboratory environment and in field trials, confirming how the platform represents a good opportunity for the testing of C-ITS applications. In our future work, our platform will be used to provide connectivity to a connected and autonomous mini-car, to perform scaled field tests of V2X-based solutions for cooperative maneuvers.

\section*{Acknowledgment} This work was partially supported by the European Union under the Italian National Recovery and Resilience Plan (NRRP) of NextGenerationEU, partnership on ``Telecommunications of the Future'' (PE00000001 - RESTART), project MoVeOver, and national research centre on mobility (CN00000023 - MOST).


\bibliographystyle{IEEEtran}  
\bibliography{biblio}

\end{document}